\documentclass[a4paper,fleqn,usenatbib]{mnras}

\usepackage{newtxtext,newtxmath}

\usepackage[T1]{fontenc}
\usepackage{ae,aecompl}

\usepackage{graphicx}	
\usepackage{amsmath}	
\usepackage{amssymb}	

\usepackage{longtable, verbatim}
\usepackage{booktabs}

\newcommand{\um}{\,\mu{\rm m}}

\newcommand{\pc}{\,{\rm pc}}
\newcommand{\degree}{^\circ}

\newcommand{\AU}{\,{\rm AU}}

\newcommand{\CL}{CL$\,07$}

\title[Polarized emission of aligned aspherical grains]{3D radiative transfer of intrinsically polarized dust emission based on aligned aspherical grains}

\author[G. H.-M. Bertrang et al.]{
G. H.-M. Bertrang,$^{1,2,3,4}$\thanks{E-mail: bertrang@das.uchile.cl}
S. Wolf$^{1}$
\\
$^{1}$Kiel University, Institute of Theoretical Physics and Astrophysics, Leibnizstr. 15, 24118 Kiel, Germany\\
$^{2}$Universidad de Chile, Departamento de Astronom\'ia, Casilla 36-D, Santiago, Chile\\
$^{3}$Universidad Diego Portales, Facultad de Ingenier\'ia, Av. Ej\'ercito 441, Santiago, Chile\\
$^{4}$Millennium Nucleus Protoplanetary Disks in ALMA Early Science, Universidad de Chile, Casilla 36-D, Santiago, Chile
}

\date{Accepted 2017 May 1. Received 2017 April 28; in original form 2016 June 7}

\pubyear{2015}

\begin{document}
\label{firstpage}
\pagerange{\pageref{firstpage}--\pageref{lastpage}}
\maketitle

\begin{abstract}
(Sub-)Millimeter observations of the polarized emission of aligned 
aspherical dust grains enable us to study the magnetic fields within 
protoplanetary disk.  However,  the interpretation of these observations is complex. One must consider the various effects that alter the measured polarized signal, such as the shape of dust grains, the efficiency of grain alignment, the magnetic 
field properties, and the projection of the signal along the line of sight. We aim at analyzing observations of the polarized dust emission by disentangling the effects on the polarization signal in the context of 3D~radiative transfer simulations.
For this purpose, we developed a code capable of simulating dust 
grain alignment of aspherical grains and intrinsical polarization of thermal dust 
emission. We find that the influence of thermal polarization and dust grain alignment 
on the polarized emission displayed as spatially resolved polarization map or as spectral energy distribution trace disk properties which are not traced in total (unpolarized) emission such as the magnetic field topology.
The radiative transfer simulations presented in this work enable the 3D analysis of intrinsically polarized dust emission -- observed with, e.g., ALMA -- which is essential to constrain magnetic field properties.

\end{abstract}



\section{Introduction}

Various physical processes in protoplanetary disks are affected by magnetic fields.
Magnetic fields influence the transport of dust and gas 
\citep[e.g.,][]{2007ApJ...654L.159C, 2014prpl.conf..411T}, the disk
chemistry \citep[e.g.,][]{2011ApJS..196...25S}, and the migration of 
plane\-tesimals within the disk \citep[e.g.,][]{2010A&A...515A..70D}. Even more importantly, magnetohydrodynamic (MHD) turbulence can 
provide the source of viscosity that expedites the accretion 
\citep{1974MNRAS.168..603L}, and thus, the 
evolution of the disk \citep{1973A&A....24..337S}. One of the most promising
mechanisms for driving turbulence, respectively accretion is the 
magneto-rotational instability \citep[MRI;][]{1991ApJ...376..214B, 
1998RvMP...70....1B}. Turbulence in unmagnetized disks is unable to
redistribute angular momentum in the disk sufficiently effective, thus, magnetic fields are 
needed to enable Shakura-Sunyaev viscosity \citep{1996ApJ...467...76B}. The 
ionization fraction in the disk is high enough for magnetic coupling of 
material over large parts of the disk especially the thermally ionized inner 
disk and the highly ionized outer disk parts \citep[e.g.,
][]{2007ApJ...659..729T,2015ApJ...799..204C, 2015A&A...574A..68F}. Yet, 
constraining the magnetic field properties of the disk observationally, its strength and 
structure, is still at the beginning. First observations of 
very young, embedded protoplanetary disks performed with SMA, CARMA, and VLA show polarized signals 
which can be interpreted as indications for toroidal magnetic field  structures
\citep{2014ApJ...780L...6R, 
2014Natur.514..597S, 2015ApJ...798L...2S}. Despite of non-detections found in more evolved protoplanetary disks performed with the SMA \citep{2009ApJ...704.1204H, 2013AJ....145..115H}, recent ALMA observations revealed the polarization strength and structure of the protoplanetary disk HD142527 on a hitherto unreached spatial resolution \citep{2016arXiv161006318K}. 

Spatially 
resolved observations of polarized millimeter continuum emission of 
aligned dust grains are best suited to reveal the magnetic field strength and structure in the protoplanetary disk \citep[e.g.,][]{2000prpl.conf..247W,
2007ApJ...669.1085C}. However, the polarization signal is strongly influenced by many 
factors, e.g., dust grain shape, dust grain alignment efficiency, magnetic 
field properties, the spatial resolution of the observation, the projection along 
the line of sight, and the influence of self-scattered thermal emission \citep[][]{2015ApJ...809...78K, 2016arXiv161006318K, 2016MNRAS.456.2794Y}. To take these effects into account, radiative 
transfer simulations are needed as analysis tools for intensive interpretation and 
prediction of observational results. A first approach has been undertaken by \cite{2007ApJ...669.1085C}, henceforth \CL. 
 \CL~employ a two-layered model of a flared circumstellar disk based on \cite{2001ApJ...547.1077C}. In this model, the surface layer contains small dust grains, while the interior of the disk is filled with large grains. Each layer has its own, constant temperature, possible photon paths are limited, and the radiation field has been fixed to perfect anisotropy. Based on this simplified model, \CL~predict that multi-wavelength infrared polarimetric studies of circumstellar disks can provide details on their magnetic field structure. Motivated by their findings, we aim at a more sophisticated treatment of aligned aspherical dust grains in protoplanetary disks and their intrinsically polarized thermal emission. We present here a 3D~radiative 
transfer study considering the effects of polarized dust emission and dust 
grain alignment of aspherical grains that has been performed with an extended version of 
the established radiative transfer code MC3D 
\citep{1999A&A...349..839W, 2003CoPhC.150...99W}.

This paper is organized as follows. The alignment of aspherical grains and implemented approach of polarized thermal dust emission are 
introduced in Sect.~\ref{align} and Sect.~\ref{thpol}.  The simulation software is introduced in Sect.~\ref{tests}. Applications of this study are shown in Sect.~\ref{app}. A final 
discussion and conclusions are given in Sect.~\ref{conc}.

\section{Alignment of aspherical grains}\label{align}
Within protoplanetary disks, dust grains are assumed to be aligned along the magnetic field by 
radiative torques \citep[e.g.,][]{Lazarian+2015}. In the presence of an 
anisotropic radiation field, aspherical dust grains, which have different 
cross sections for left- and right-handed polarized light, get aligned with their 
longest grain axis perpendicular to the magnetic field  
\citep[][]{1976Ap&SS..43..291D, 2007JQSRT.106..225L}. Additional grain 
alignment mechanisms, i.e., paramagnetic alignment and mechanical alignment by 
either super\-sonic or subsonic flows, are negligible in protoplanetary disks: 
Paramagnetic alignment \citep{1951ApJ...114..206D} affects only dust grains 
smaller than $0.1\,\mu{\rm m}$ \citep{Lazarian+2015, 2015ARA&A..53..501A}. Mechanical alignment by 
supersonic flows \citep{1969Phy....41..100P} does not occur because of purely 
widely observed subsonic velocities, \citep[e.g.,][]{2011ApJ...727...85H} and 
because of the subsonic nature of MRI turbulence \citep{2011ApJ...735..122F}. 
And mechanical alignment by subsonic flows \citep{2007ApJ...669L..77L} appears 
to be inefficient in general, since aspherical grains may not show well-defined
helicity in the process of grain interaction with gaseous flows 
\citep{Lazarian+2015, 2015ARA&A..53..501A}. In brief, the usual picture of aspherical grains 
aligned perpendicular to magnetic field lines applies. The polarized thermal 
emission of these grains traces the magnetic field within a protoplanetary disk. 

We implement radiative torques in the 3D~radiative transfer code MC3D on the basis of the approach outlined by \CL. This approach allows us to consider the effect of radiative torques in a sophisticated disk model, taking 
the anisotropy of the radiation field into account. The mathematical description of radiative torques 
used in this implementation as well as in \CL~goes back to the fundamental 
work of \cite{1996ApJ...470..551D}.

Dust grains in a protoplanetary disk are exposed to a radiation field, consisting of the emission of the central star and the emission of the dust itself. Thus, the grain is subject to a steady radiative torque. For the alignment of a grain, it is necessary that its rotational energy exceeds pure thermal motion. This ratio is commonly expressed by the factor \citep{1996ApJ...470..551D}:

\begin{align}
 \left(\frac{\omega_{\rm rad}}{\omega_{\rm T}}\right)^2  =&\,\frac{5\alpha_{\rm 1}}{192\delta^2}\left(\frac{u_{\rm rad}}{n_{\rm H}k_{\rm B}T}\right)^2\left(\frac{\rho a_{\rm eff} \lambda^2}{m_{\rm H}}\right)\,\cdot\tag*{}\\
 &\cdot\,\gamma_{\rm rad}^2\left<\mathbf{Q}_{\rm \Gamma}\left(\theta,\varphi\right)\right>^2\cdot\mathbf{a}_{\rm 1}^2\left(\frac{\tau_{\rm drag}}{\tau_{\rm drag, 
gas}}\right)^2\\
 =&\,4.72\times 10^9\,\cdot\,\frac{\alpha_{\rm 1}}{\delta^2} \rho_{\rm 3} a_{\rm -5}\,\cdot\tag*{}\\
 &\cdot\, \left(\frac{\gamma_{\rm rad}\cdot  u_{\rm rad}}{n_{\rm H}k_{\rm B}T}\right)^2 \left(\frac{\lambda}{1\um}\right)^2 Q_{\rm \Gamma}^2\left(\frac{\tau_{\rm  drag}}{\tau_{\rm drag, sgas}}\right)^2\,,
\label{eq:omega}
\end{align}

\noindent where $\omega_{\rm rad}$ is the angular frequency due to radiative torques, $\omega_{\rm T}$  the thermal angular frequency, which is the 
rate at which the rotational kinetic energy of the grain is equal to $k_{\rm 
B}T/2$, $Q_{\Gamma} = \left<\mathbf{Q}_{\rm 
 \Gamma}\left(\theta,\varphi\right)\right>\cdot\mathbf{a}_{\rm 
1}$ is the efficiency factor of radiative torques, the geometrical parameters 
$\delta\approx 2$, $\alpha_{\rm 1}\approx 1.745$ for a grain with axis ratio 
$2$:$2$:$1$, gas mass density  $\rho_{\rm 3} = \rho/3\,{\rm g\,cm^{-3}}$ 
(assuming a gas-to-dust ratio of 100), and $a_{\rm -5}= a/10^{-5} {\rm cm}$ is 
the equivolume grain radius. The degree of anisotropy
$\gamma_{\rm rad}$ is defined by

\begin{equation}
 \gamma_{\rm rad} = \frac{\left|\int_{\Omega} I_{\rm 
\lambda}\left(\mathbf{k}\right)\mathbf{k}\, {\rm d}\Omega \right|}{\int_{\Omega}
I_{\rm \lambda}\left(\mathbf{k}\right) {\rm d}\Omega}\,,
\label{eq:anisogamma}
\end{equation}

\noindent with the wave vector $\mathbf{k}$, the wavelength-dependent intensity 
$I_{\rm \lambda}$, and the solid angle~$\Omega$ \citep{2007ApJ...663.1055B}. The degree of anisotropy lies in the range of $\left[0,\,1\right]$, where $0$ corresponds to an isotropic radiation field and $1$ to an unidirectional radiation field. The here presented temperature and anisotropy simulations are conducted self-consistently over a wavelength spectrum from $0.5\,\umu m$ to $1000\,\umu m$. 

\noindent The rotational damping time \citep{1996ApJ...470..551D},

\begin{equation}
\tau_{\rm drag}^{-1} = \tau_{\rm drag, gas}^{-1} + \tau_{\rm drag, em}^{-1}\,,
\end{equation}

\noindent is composed of (i) rotational damping caused by gas-dust collisions,

\begin{equation}
 \tau_{\rm drag, gas} = \frac{\pi\alpha_{\rm 1}\rho a_{\rm eff}}{3\delta n_{\rm 
H}\left(2\pi m_{\rm
H}k_{\rm B}T\right)^{1/2}}\,,
\label{eq:taudraggas}
\end{equation}

\noindent and of (ii) 
rotational damping caused by absorption and emission of photons by the grain 
\citep[][]{1971ApJ...167...31P, 1993ApJ...418..287R}. Assuming that the grain is 
heated by the star to a tempe\-rature $T_{\rm dust}$, the damping time due 
to thermal emission may be written as

\begin{equation}
 \tau_{\rm drag, em} = \frac{8\,\alpha_{\rm
1}\left(\beta+3\right)}{5}\,\frac{\zeta\left(\beta+4\right)}{
\zeta\left(\beta+3\right)}\,\frac{\rho a^3_{\rm
eff}\left(k_{\rm B}T_{\rm dust}\right)^2}{\hbar c u_{\rm rad} \left<Q_{\rm 
abs}\right>}\,.
\label{eq:taudragem}
\end{equation}

\noindent The quantity $\zeta(x)$ represents the Riemann $\zeta$-function, and the parameter 
$\beta=2$ which is a constant of order unity and depends on the character of 
the scattering at the boundary \citep{1984ApJ...285...89D, 
1994ASPC...58..227D}. However, the rotational damping time $\tau_{\rm drag, em}$ 
is insensitive to the exact value of $\beta$ \citep{1996ApJ...470..551D}. In 
Eq.~(\ref{eq:taudragem}) it is assumed that emitted photons have an angular
momentum $\hbar$ relative to the grain center of mass which is true for $a\ll 
hc/k_{\rm B}T_{\rm dust}=800\,(18{\rm K}/T_{\rm dust})\um$ and fulfilled by 
typical conditions in protoplanetary disks 
\citep[e.g.,][]{2002ocd..conf....1V}. The additional damping caused by 
absorption of photons is smaller than the damping time due to thermal emission 
by a factor $\sim T_{\rm dust}/T_{\rm rad}\approx 1/500$, where $T_{\rm rad}$ 
is the color temperature of the dust heating radiation 
\citep{1979ApJ...231..404P}, and thus, it is neglected. However,
the calculations of this work and of \CL~ show that $(\tau_{\rm 
drag}/\tau_{\rm drag, gas})^2 \approx 1$ is valid throughout in the disk.

All quantities in Eq.~(\ref{eq:omega}) can be extracted from radiative transfer 
simulations or are fixed 
parameters, except for the efficiency factor of radiative torques, 
$Q_{\Gamma}$. The computation of $Q_{\Gamma}$ for aspherical dust grains are 
performed with the DDSCAT software package \citep[]{1994JOSAA..11.1491D, 2004astro.ph..9262D, 1996ApJ...470..551D}  by \cite{2005ApJ...631..361C} and \cite{2007ApJ...669L..77L} for 
grains with equivolume radii between $0.1\um$ and $100\um$. However, in 
protoplanetary disks dust grains grow and have sizes of about 
$1000\um$. For the resulting $\lambda/a$ ratios, the computational time of 
DDSCAT increases significantly.  For this reason, it is more applicable to 
approximate $Q_{\Gamma}$ (see Fig.~2 in \CL) with

\begin{equation}
 Q_{\Gamma} = \begin{cases}
    \sim \mathcal{O}(1) & \text{, if } \lambda\sim a\,,\\
    \sim (\lambda / a)^{-1/3} & \text{, if } \lambda > a.
 \end{cases}
\label{eq:Qgama}
\end{equation}

\noindent Together with Eq.~(\ref{eq:Qgama}), Eq.~(\ref{eq:omega}) may be 
written  for $\lambda > a$ as

\begin{equation}
 \left(\frac{\omega_{\rm rad}}{\omega_{\rm T}}\right)^2 \approx  
\left(\frac{\omega_{\rm rad}}{\omega_{\rm
T}}\right)^2_{\rm \lambda\sim a} \left(\frac{Q_{\rm\Gamma, \lambda\sim 
a}}{Q_{\rm\Gamma, \lambda}}\right)^2 \approx
\left(\frac{\omega_{\rm rad}}{\omega_{\rm T}}\right)^2_{\rm \lambda\sim a}
\left(\frac{\lambda}{a}\right)^{-6}\,,
\label{eq:implemented}
\end{equation}

\noindent where

\begin{align}
 \left(\frac{\omega_{\rm rad}}{\omega_{\rm T}}\right)^2_{\rm \lambda\sim a} 
\approx&\,4.72\,\times\,10^9\,\frac{\alpha_{\rm 1}}{\delta^2}\,\cdot\tag*{}\\
&\cdot\, \rho_{\rm 3}a_{\rm 
-5}\left(\frac{\gamma_{\rm rad} u_{\rm
rad}}{n_{\rm H}k_{\rm 
B}T}\right)^2\,\left(\frac{\lambda}{1\um}\right)^2\,\left(\frac{\tau_{\rm 
drag}}{\tau_{\rm drag, gas}}\right)^2.
\end{align}

\noindent Grains with rotational energies much higher compared to thermal 
energy 
will stably align with the magnetic field. This condition is satisfied by 
$(\omega_{\rm rad}/\omega_{\rm T})^2 > 10$, as shown in the stability study of 
\cite{1997ApJ...480..633D}. It is assumed that all grains in the current cell 
are perfectly aligned as soon as  $(\omega_{\rm rad}/\omega_{\rm T})^2 > 10$.

\mbox{ }

\begin{figure}
\includegraphics[width=\columnwidth, 
keepaspectratio]{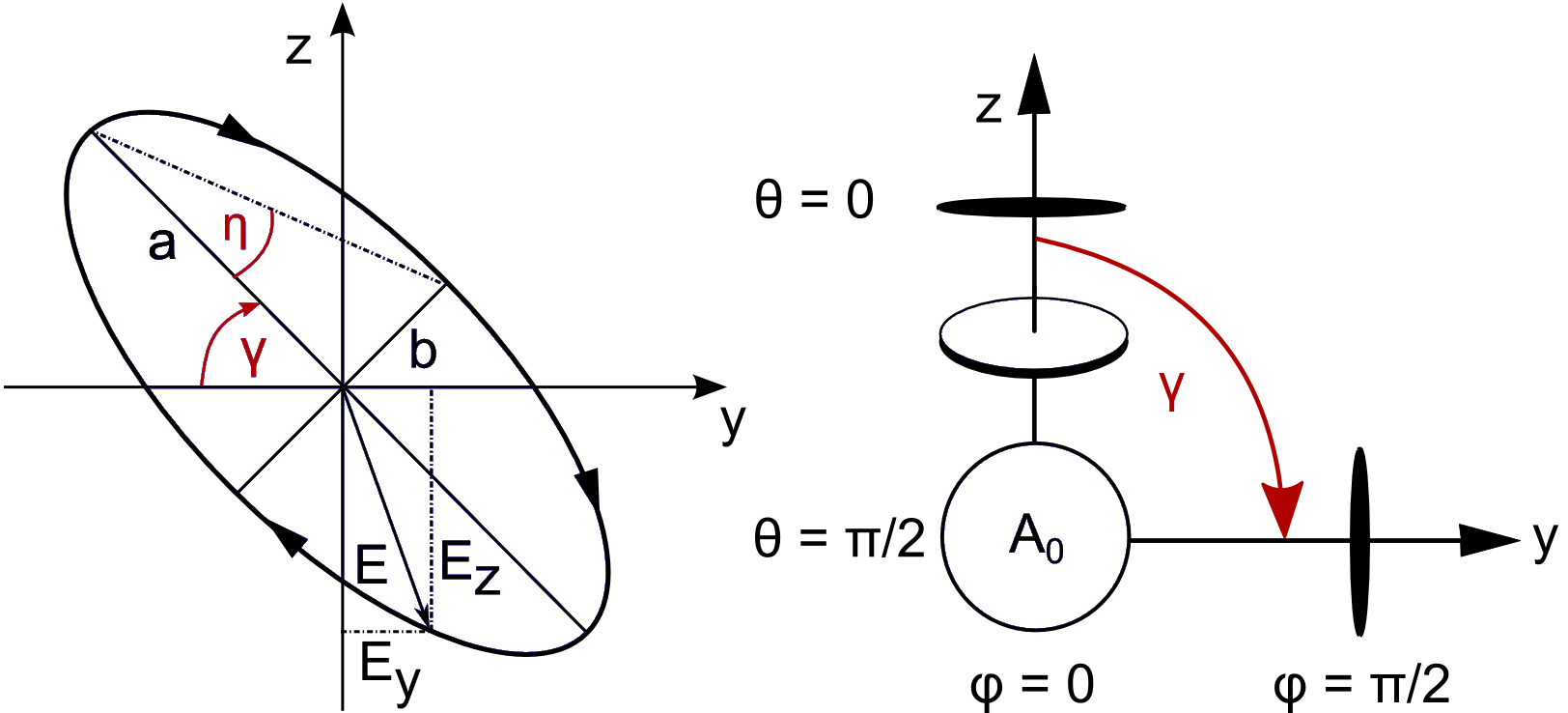}
\caption{\textsc{Angle definitions for thermal polarization:} {\it
Left:} The polarization ellipse of a light wave 
$\mathbf{E}$.
 {\it Right:} The position and orientation of the grain can be defined 
relative to the observer who is, by definition, observing along the $x$-axis.}
\label{pic:thermpol_definitions}
\end{figure}

\section{Intrinsic thermal polarization} \label{thpol}
The intrinsic thermal polarization depends on the shape of the dust grains as well as on their relative orientation along the line-of-sight to the observer. The latter is used to trace the magnetic field structure. In this section, we outline the calculation of the instrinsic thermal polarization for individual grains. This provides the basis for the 3D~radiative transfer simulations of the net line-of-sight polarization (see Sect.~\ref{tests})

The quantification of polarization of a light wave is given by 
characteristic intensities, the Stokes parameters, $I, Q, U, V$ 
\citep{1852TCaPS...9..399S}:

\begin{equation}
     \begin{aligned}
    I &\quad = \quad I\,, \\
    Q &\quad = \quad I_{\rm P}\cos2\eta\cos2\gamma\,,\\
    U &\quad = \quad I_{\rm P}\cos2\eta\sin2\gamma\,,\\
    V &\quad = \quad I_{\rm P}\sin2\eta\,.
    \end{aligned}
 \label{eq:stokesparameters}
\end{equation}

\noindent where $I$ represents the total (unpolarized) intensity, $Q$ and 
$U$ characterize the linear polarization, while $V$ represents circular 
polarization. The factor of two in front of the polarization angle, $\gamma$, 
is due to the fact that any polarization ellipse (see 
Fig.~\ref{pic:thermpol_definitions}) is indistinguishable from one 
rotated by $180\degree$. A polarization ellipse is another useful concept to characterize the polarization state of a light wave 
$\mathbf{E}$. It is defined by 
its direction of rotation, by its inclination relative to a reference axis, 
the polarization angle, $\gamma$, by the relation
of its half-axes, $\eta$, and its radiant 
intensity, $c^2 = a^2 + b^2$. The opening angle 
$\eta$ can also be expressed in the observer's frame by 
$\theta$ and $\varphi$ as described in Eq.~(\ref{eq:eta2}). The factor of two in front of the opening angle 
of the polarization ellipse, $\eta$, in Eq.~(\ref{eq:stokesparameters}) indicates that an ellipse 
is indistinguishable from one with the semi-axis lengths swapped accompanied by 
a $90\degree$ rotation.

Aspherical dust grains emit intrinsically polarized waves due to their 
aspherical shape. The relative orientation of the grain and 
the observer determines the polarization of the detected wave, 
characterized by the angles $\gamma$, with
\begin{equation}
\gamma =
\frac{1}{2}\arctan\left(\frac{U}{Q}\right)\,,\qquad\gamma\in\left[0,
\pi\right]
\label{eq:polang}
\end{equation}
\noindent and $\eta$ that depends on the orientation of the grain relative to 
the orientation of the observer. By definition, the observer is observing along the $x$-axis of a Cartesian system. The polarization 
angle, $\gamma$, and the opening angle of the ellipse, $\eta$, can be derived 
from the angles $\theta$ (in the xz-plane) and $\varphi$ (in the xy-plane) within the observer's frame  (see Fig.~\ref{pic:thermpol_definitions}). The
projected area of the grain surface is mini\-mized for $\theta = 0$ or $\varphi = 
\pi / 2$, and is maximized for $\theta = \pi / 2$
or $\varphi = 0$. For the purposes of this description it is assumed that the grain is of oblate 
shape with infinitesimal width in one direction as sketched in Fig.~\ref{pic:thermpol_definitions}.  However, the behaviour of the relative change in polarization degree is not affected by this approach.  We adjust the simulation to more realistic grain shapes, by adapting the polarization ellipse accordingly to the semi-axes' parameters $(a,b)$ of a given grain shape.
In general, the 
aspherical, irregularly shaped grain must rotate rapidly to be aligned with the 
magnetic field. Thus, it can be approximated by an ellipsoidal shape. The 
opening angle $\eta$ is defined by the projected surface of the grain, 
$A(\theta, \varphi)$. The minimal projected grain surface, resp. the maximal possible polarization degree is given by the ratio of the grain axes which is a free parameter. Following 
 Fig.~(\ref{pic:thermpol_definitions}), $\,A(\theta, \varphi)$ can be described 
by
\begin{equation}
  A(\theta)  = A_{\rm 0} \left|\sin\theta\right|,\quad  A(\varphi) = A_{\rm 
0} \left|\cos\varphi\right|,
\end{equation}
where $ A_{\rm 0}$ is the maximum of the projected surface. Thus, $A(\theta, 
\varphi)$ reads
\begin{equation}
  A(\theta, \varphi) = A_{\rm 0} \left|\sin\theta\cdot\cos\varphi\right|.
\end{equation}

\noindent For any symmetric grain, the observed length of its longest grain 
axis, $a$, is independent of the observer's position, thus
\begin{equation}
 A_{\rm 0} = \pi a^2.
\end{equation}

\noindent Together with the definition of $\eta$, 
\begin{equation}
\tan\eta = \frac{b}{a}, \qquad\qquad 
\eta\in\left[-\frac{\pi}{4},\frac{\pi}{4}\right],
\label{eq:eta}
\end{equation}
\noindent where $a$ and $b$ are the half-axes of the polarization ellipse, and 
the common definition of the area of an
ellipse, this results in the opening angle defined by the observer's position:
\begin{equation}
 \eta\left(\theta, \varphi\right) = 
\arctan\left|\sin\theta\cdot\cos\varphi\right|.
\label{eq:eta2}
\end{equation}

\noindent Then, the polarization state of the light wave $\mathbf{E}$ is given by 
the degree of polarization, $P$, and the polarization angle, $\gamma$ 
(Eq.~\ref{eq:polang}). $P$ can be splitted into a linearly polarized part, 
$P_{\rm L}$, as well as in a circularly polarized part, $P_{\rm C}$, with
\begin{equation}
 P_{\rm L} = \frac{\sqrt{Q^2+U^2}}{I}\,, \qquad\qquad P_{\rm C} = \frac{V}{I}\,,
\end{equation}

\noindent Since the magnetic field structure is traced by linear 
polarization only, we consider only $P_{\rm L}$ to which we refer to as $P$ in 
the rest of this paper.

\section{The simulation software}\label{tests}
This implementation of radiative torques and polarized thermal emission of 
aspherical dust grains into MC3D has undergone extensive testing. Analytical 
tests show the high precision of the thermal polarization computations which 
are accurate down to the numerical limits. However, due to its dependence on the radiation field, the dust 
grain alignment must be tested numerically, which is presented 
in Sect.~\ref{sec:test_sphere}. Limitations of this approach are discussed in Sect.~\ref{sec:limits}.

\begin{table}
\resizebox{\columnwidth}{!}{%
\hspace{-1.5em}                       
\begin{tabular}{c c c c c c c}\toprule\toprule
$R_{\rm in}$ & $R_{\rm out}$ & Distance & $a_{\rm dust}$ & $M_{\rm dust}$ & 
$R_{\rm *}$ & $T_{\rm *}$ \\

$\left[\AU\right]$ & $\left[\AU\right]$ & $\left[\pc\right]$ & 
$\left[\um\right]$ & $\left[M_{\rm\sun}\right]$ & $\left[R_{\rm\sun}\right]$ & 
$\left[T_{\rm\sun}\right]$\\\midrule

$1$ & $200$ & $100$ & $[0.01, 1000]$ & $10^{-6}$ & 
$2.5$ & $4000$ \\\bottomrule
\end{tabular}}
\caption{Model parameters of the test case \emph{sphere}.}
\label{tab:test_sphere2} 

\end{table}

\subsection{Test case: Spherical density distribution}\label{sec:test_sphere}

\begin{figure*}
\includegraphics[width=0.3\linewidth]{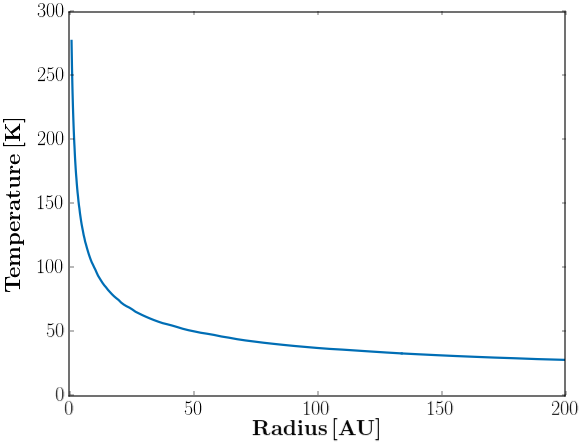}
\includegraphics[width=0.3\linewidth]{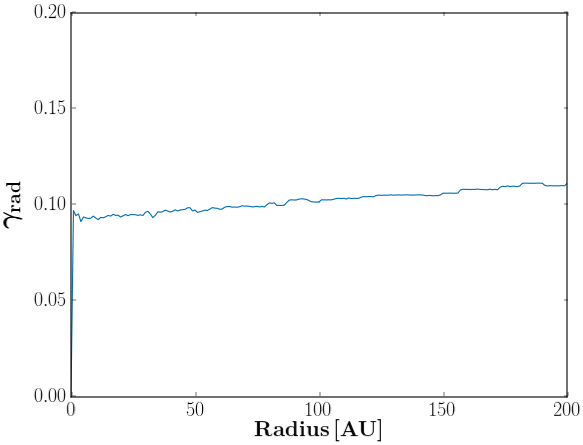}
\includegraphics[width=0.3\linewidth]{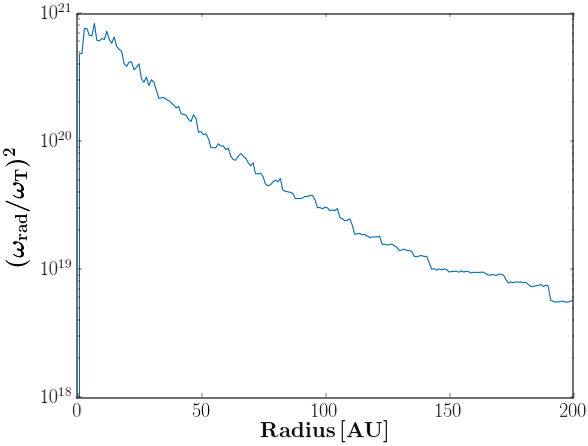}

\caption{ {\sc A spherical dust density distribution} in temperature $T$ 
(\emph{left}), degree of anisotropy of the radiation field
 $\gamma_{\rm rad}$ (\emph{center}), and dust grain alignment efficiency
$(\omega_{rad}/\omega_{T})^2$ (\emph{right}).}
\label{pic:test_sphere}
\end{figure*}

To test the implementation of the dust grain alignment we choose a spherical dust density distribution, 
$\rho\left(r\right) \sim r^{-1}$, with further model parameters 
described in Tab.~\ref{tab:test_sphere2}. For this model, distributions of 
temperature, anisotropy of the radiation field, and dust grain alignment have 
been computed self-consistently (see Fig.~\ref{pic:test_sphere}). The numerical grid of this 1-dimensional test problem is discretized in  $N_{\rm r} = 256$ cells on which the distributions of temperature, the degree of anisotropy of the radiation field, and grain alignment efficiency are computed at $45$~discrete logarithmically distributed wavelengths within $\left[1\um,1000\um\right]$ with $10^7$ photons. 

The temperature distribution, $T\left(r\right)$,  of a 
spherical density distribution heated only by a central star shows the expected radial decrease. 
Close to the central star where dust density and stellar radiation are 
have reached their maximum values, the temperature reaches its maximum, as well. Towards the boundaries of the model space, it is decreasing radially. 

The anisotropy of the radiation field, $\gamma_{\rm rad}\left(r\right)$, is slightly increasing radially. Being located close to the central star its isotropic reemission of high-energy photons results in an anisotropy of the net radiation field. At 
radii with decreasing temperatures, the dust emits less high-energy photons,  while photons from 
the central regions are efficiently shielded by absorption of the more dense inner regions. In the 
boundary cells, the anisotropy of the radiation field reaches its maximum.
Due to the negligence of an external radiation field, photons are only able 
to enter these cells from more inner regions.

The alignment efficiency of dust grains, $(\omega_{rad}/\omega_{T})^2$, is very 
sensitive to the ratio of temperature, energy density of the radiation 
field, and degree of anisotropy of the radiation field, as well as to the local mass 
density (see Eq.~\ref{eq:implemented}). In general, 
$(\omega_{rad}/\omega_{T})^2$ exceeds $10$ throughout the entire model space, resulting in 
an alignment of all grains in this test case. As expected for the given model, 
the value of $(\omega_{rad}/\omega_{T})^2$ decreases radially with temperature, energy density, and mass density. 

\subsection{Limitations of this implementation}\label{sec:limits}
Although the polarization of dust emission of aligned 
dust grains has been implemented successfully, there are limitations to this approach that are discussed here:

{\it i)$\,\,$Grain alignment:}\indent
In the description of grain alignment we assume that all grains within a given cell are aligned perfectly as soon 
as $(\omega_{\rm rad}/\omega_{\rm T})^2 > 10$ \citep[][]{1996ApJ...470..551D, 1997ApJ...480..633D}.

{\it ii)$\,\,$Scattering:}\indent For the implementation of 
aspherical dust grains into MC3D, we apply a hybrid approach regarding the dust grain shape. During the self-consistent calculations of temperature 
distribution and anisotropy of the radiation field, scattering effects under consideration of both a dust grain size distribution and a broad wavelength spectrum are 
indispensable. Scattering by spherical grain shapes is relatively easy to handle 
numerically (e.g., Mie scattering). However, scattering by aspherical grain 
shapes over a broad range of grain sizes and wavelengths is still an issue for the mathematical description and numerical implementation \citep{2004ASPRv..12....1V}. For that reason, MC3D 
processes the photons through the model space assuming spherical grain shapes 
during the computation of dust temperature distributions and anisotropy of the 
radiation field, and skips to aspherical grains for calculating the 
polarization state. Here, the scattering is neglected, and thus, the 
significance of these simulations is limited to long wavelengths where 
scattering effects play only a tangential role for observations of 
protoplanetary disks. However, this study already opens unprecedented 
possibilities to investigate the effects of polarized dust emission and grain 
alignment in protoplanetary disks. 

\section{Applications} \label{app}

This 3D radiative transfer study, now including polarized dust emission of 
aspherical grains, aims at modeling, character\-izing, and explaining 
polarimetric observations of T-Tauri disks. Applications of these 
new features are discussed here with the goal of comparing the results of this study with the findings described in the study of \CL. Note that, due to the partly significant differences of both approaches (the work of \CL~assumes, e.g., isothermality of the disk interior, limitations of possible photon paths, and neglection of the orientation of polarization vectors in SED simulations)  the evaluation can only be done qualitatively. Yet, due to the lack 
of other work on thermal polarization simulations of aligned dust 
grains in protoplanetary disks, this method is the most suitable.

\begin{figure}
\centering \includegraphics[width=0.9\columnwidth, %
keepaspectratio]{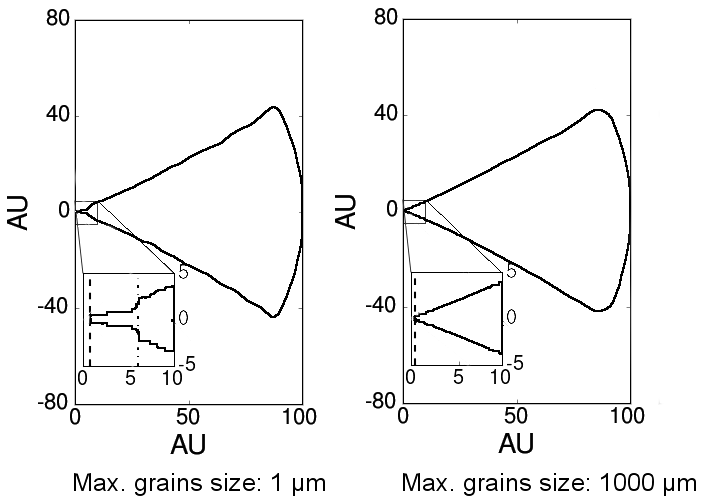}
 \caption[\textsc{Grain alignment} in the disk.]{\textsc{Grain alignment in the 
disk} (shown perpendicularly to the disk 
mid-plane) with maximum grain sizes of $1\um$ \emph{(left)} and $1000\um$ 
\emph{(right)}. Within the black contour lines, the grain alignment efficiency exceeds the threshold of 10, i.e., the grains are aligned. Small grains start to align
close to the mid-plane at $r\geq0.7\AU$ ({\sl dashed line}), at $r\geq6\AU$ 
({\sl dashed-dotted line}) grains also align towards the
disk surface. Large grains with radii~$\leq1000\um$ have only one initiation
of grain alignment at $r = 0.08\AU$. These findings correspond very well to the findings of \CL~(Fig.~3, Fig.~4; $r\gtrsim0.7\AU$ in the surface layer and $r\gtrsim0.05\AU$ in the disk interior).
}\label{pic:align_comparison}
\end{figure}

In both studies, the magnetic field is 
assumed to be regular and toroidal, and both assume a flared disk structure. This study uses the approach of \cite{1973A&A....24..337S},

\begin{equation}
\rho\left(R,\,z\right) = \rho_{\rm 0} \left(\frac{R}{r_{\rm 0}}\right)^{-\alpha}\exp\left(-\frac{1}{2}\left[\frac{z}{h\left(R\right)}\right]^2\right),~ h\left(R\right) = h_{0}\left(\frac{R}{r_{0}}\right)^{\beta}
\end{equation}

\noindent with $\alpha=1.2,\,\beta=1.14$ and with the same 
power-law for the grain size distribution as applied by \CL,  $dN \propto 
a^{-3.5} {\rm d}a$ \citep{1977ApJ...217..425M}. We assume a grain composition of 
a mix of $62.5\%$~astronomical silicate and $37.5\%$~graphite with a total bulk density of $2.7\,g\,cm^{-3}$ and a continual grain size distribution \citep[Tab.~\ref{tab:mc3d_testinput};][]{2001ApJ...548..296W}, while \CL~ consider only 
astronomical silicate, with and without water ice mantle, arranged in a thin surface layer ($a_{\rm max}=1\um$) and the disk interior ($a_{\rm max}=1000\um$). 
Furthermore, this study computes  the temperature distribution of the disk and the anisotropy of the radiation field self-consistently 
based on the Monte-Carlo radiative transfer method and assumes local 
thermodynamic equilibrium in each grid cell only. \CL~ limit the radiation flux from the surface layer and the disk interior to a narrow spectrum around the wavelength $\sim 3000/T_{\rm 
ds,i}\um$, limit the direction of the radiation flow within the disk to the vertical 
direction only, and assume that the disk interior is isothermal.
Both models neglect effects of polarizing scattering. Indeed, effects of scattering are less important in the 
sub-millimeter/millimeter wavelength regime.

\begin{figure}
\centering
\includegraphics[width=0.8\columnwidth, %
keepaspectratio]{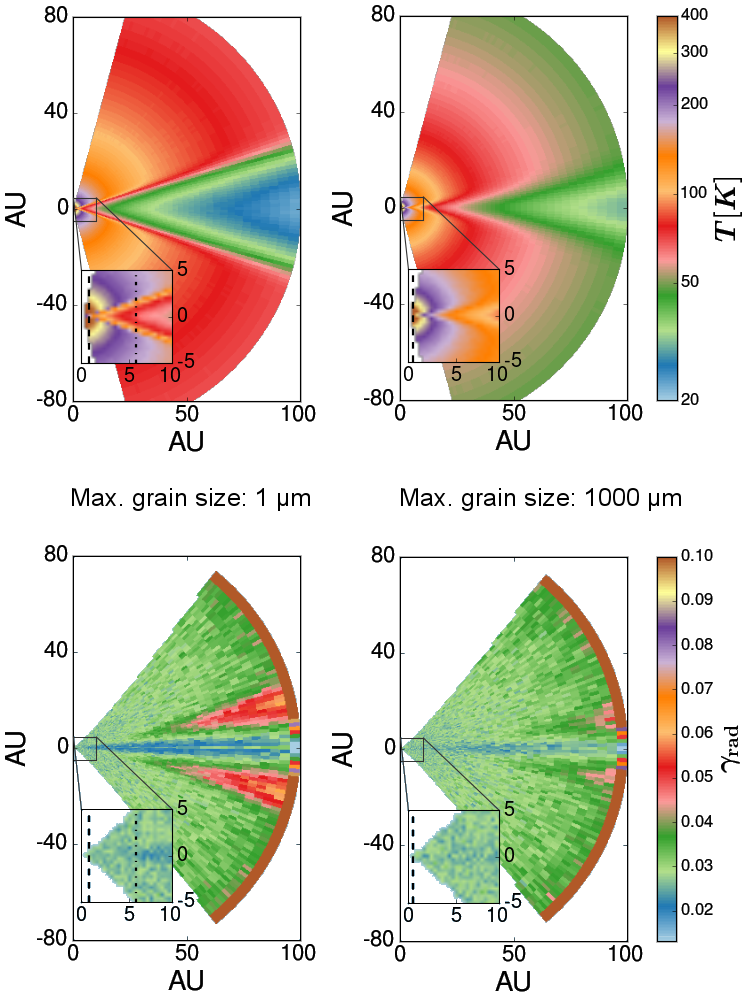}
 \caption[\textsc{Temperature distribution and anisotropy of the radiation 
field} in the disk.]{\textsc{Temperature distribution and anisotropy of the 
radiation field} in the disk. \emph{Top:} The temperature distribution of the 
disk perpendicular to its mid-plane for $a_{\rm max}=1\um$ (\emph{top left}) 
and $a_{\rm max}=1000\um$ (\emph{top right}). \emph{Bottom:} The anisotropy of
the radiation field in the case of $a_{\rm max}=1\um$ (\emph{bottom left}) and 
$a_{\rm max}=1000\um$ (\emph{bottom right}). The different maximal grain sizes 
change the optical depth of the disk. The optical thick disk (\emph{left}) 
shows sharp gradients in anisotropy and temperature, the optical thin(er) disk 
(\emph{right}) shows much smoother gradients.  While the smoother gradients 
lead to only one onset of grain alignment (dashed lines), the sharp gradients 
benefit the second jump in grain alignment (dashed-dotted 
line).}\label{pic:aniso_temp}
\end{figure}

\begin{figure*}
 \centering
\begin{minipage}{0.8\textwidth}
 \includegraphics[width=\textwidth, 
keepaspectratio]{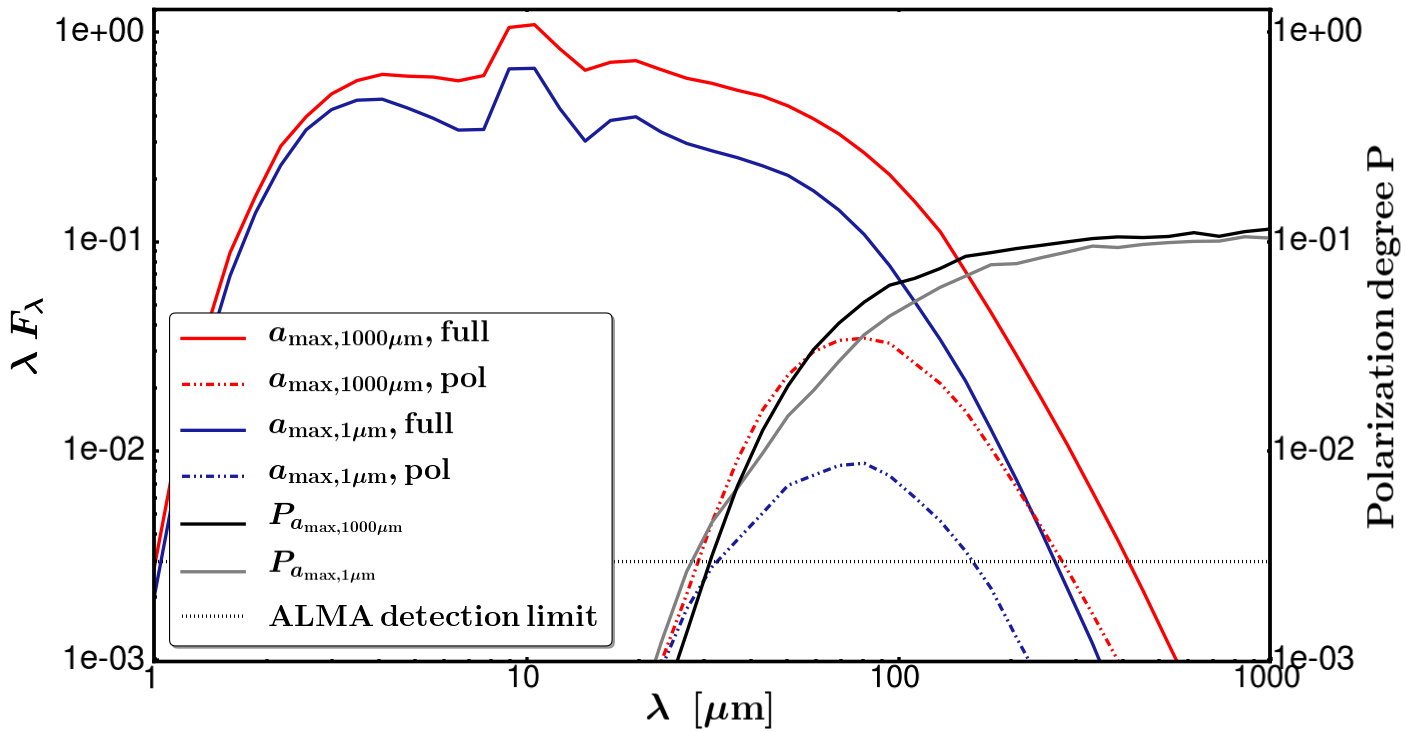}\vspace*{0.5ex}
\end{minipage}
\caption[\textsc{Spectral Energy Distribution}]{ \textsc{Spectral Energy
Distribution. The vertical axes ($\lambda F_{\rm \lambda}$ and $P$) are in arbitrary
units. Total (full) emission of the model with $a_{\rm max}=1\um$ (\emph{blue 
solid line}), polarized emission for $a_{\rm max}=1\um$ (\emph{blue dashed-dotted
line}), and total emission of the model with 
$a_{\rm max}=1000\um$ (\emph{red solid line}), polarized emission for $a_{\rm 
max}=1000\um$  (\emph{red dashed-dotted line}); all models are computed at a disk inclination of 
$i=10\degree$. The polarization degree of both models is given by the black/grey line. The detection limit for polarization of extended sources with ALMA is given by the black dotted line. Our findings correspond to the findings shown in \CL,~Fig.$\,8$, qualitatively. 
Note that \CL~ignored the direction of polarization 
vectors for these calculations. } \vspace*{-2ex}}\label{pic:sed_comp}
\end{figure*}

In order to compare both implementations to each other, we choose a parameter setup similar to the setup of parameters of \CL~(see 
Tab.~\ref{tab:mc3d_testinput}). In the following, the grain
alignment in the disk as well as the resulting SEDs and 
polarization maps are discussed. 

\begin{table}
\caption{Initial parameters for the case {\sl protoplanetary disk}.}\label{tab:mc3d_testinput}
\label{tab:test_sphere}     
\resizebox{\columnwidth}{!}{%
\hspace{-1.5em}                         
\begin{tabular}{c c c c c c c}\toprule\toprule
$R_{\rm in}$ & $R_{\rm out}$ & Distance & $a_{\rm dust}$ & $M_{\rm dust}$ & 
$R_{\rm *}$ & $T_{\rm *}$ \\

$\left[\AU\right]$ & $\left[\AU\right]$ & $\left[\pc\right]$ & 
$\left[\um\right]$ & $\left[M_{\rm\sun}\right]$ & $\left[R_{\rm\sun}\right]$ & 
$\left[T_{\rm\sun}\right]$\\\midrule

$0.02$ & $100$ & $140$ & $[0.01, 1000]$ & $10^{-4}$ & $2.5$ 
& $4000$\\\bottomrule
\end{tabular}
}
\end{table}

\subsection{Grain alignment in disks} 

The dust grain alignment perpendicular to the mid-plane of the protoplanetary 
disk is shown in Fig.~\ref{pic:align_comparison}. For this study, two simulations with 
corresponding maximal grain sizes are presented since \CL~discuss the 
surface layer ($a_{\rm max}=1\um$) and the disk interior ($a_{\rm 
max}=1000\um$) separately. We apply a grid with logarithmic spacing to spatially resolve the inner region of the disk on a level of $10^{-3}\,$AU \citep[for more details on the grid architecture, see][]{1999A&A...349..839W, 2003CoPhC.150...99W}. This study demonstrates that the alignment of 
grains is dependent on the grain size. Small grains with radii~$\leq1\um$ show 
two initialization steps of grain alignment, the first step in the mid-plane of the 
disk at  $r\approx0.7\AU$, the second step towards the surface of the  disk at $r=6\AU$. 
The first step corresponds very well to the starting point of grain 
alignment in the study of \CL. The 
grain alignment depends significantly on the ratio of temperature, energy 
density, and anisotropy of the radiation field,  as well as on the local mass 
density (see Eq.~\ref{eq:implemented}). Small grains are very poor absorbers 
and emitters of radiation, thus, the disk with $a_{\max}\leq1\um$ is optically 
thick. As a result, temperature and anisotropy of the radiation field rapidly 
decrease with distance to the center of the disk. At the initial point of the 
dust grain alignment towards the surface of the disk, at $r=6\AU$, sharp 
gradients are present in the anisotropy of the radiation field
and in the temperature distribution (see Fig.~\ref{pic:aniso_temp}) resulting 
in 
$(\omega_{\rm rad}/\omega_{\rm T})^2$ which then exceeds $10$ slightly.\\
\indent Large grains with radii~$\leq1000\um$ have one initiation of grain
alignment at $r = 0.08\AU$ what again corresponds very well to the findings of 
\CL~ who determined the initialization point for large grains in their disk interior 
to $r=0.05\AU$ (see Fig.~\ref{pic:align_comparison}). Large grains are much 
more efficient in re-emitting the absorbed radiation. As a result, they cool 
down more efficiently, are optically thin(er) compared to small grains,
and have much smoother gradients in temperature and anisotropy of the radiation 
field.

\subsection{Spectral energy distribution} 

The spectral energy distributions for polarized dust emission of 
aligned aspherical grains computed within the context of this study are shown in 
Fig.~\ref{pic:sed_comp} and compared to the SEDs in \CL,~Fig.~8.
The SEDs of both models 
show an emission peak around $10\um$ which is the characteristic peak of 
silicate \citep{1984ApJ...285...89D}. Additionally to this, the SED of \CL~ 
shows a second and third peak at $40\um$ and $60\um$ caused by the water ice mantle 
of the grains in their model. \CL~ 
ignored the orientation of polarization vectors in their calculations of the 
SEDs. The observed polarization degree is very sensitive to the orientation of 
polarization vectors, especially in the case of aligned dust grains. The new 
features of the study presented in this work consider this effect. With a disk inclination of 
$i=10\degree$, a good agreement with the calculations of \CL~was found by 
comparison of the behavior of the total emission relative to the polarized 
emission in both studies. As expected from the optical properties with respect 
to the wavelength, large dust grains dominate the emission spectrum
at long wavelengths. This is seen in the total emission and the polarized
emission in both studies. \CL~and this work are neglecting polarizing scattering 
effects. With respect to the 
model differences, the relative behavior of total to polarized emission of both 
grain size distributions of \CL~ and this work fit well to each other.

Furthermore, we find polarization degrees, $P$, for both models in the order of $\sim 10\%$ across long wavelengths (Fig.~\ref{pic:sed_comp}). At wavelengths below $100\um$, the strength of the intrinsically polarized emission drops significantly.

\subsubsection{The effect of inclination on spectro-polarimetry}\label{sec:modelsed}
Since polarimetry is very sensitive to projection effects as well as to spatial resolution,  the inclination of the disk is a mayor influence on the detected polarized signal. Because of the radial 
symmetry of the polarization pattern stemming from a toroidal
magnetic field structure if the disk is face-on oriented to the observer,  the 
polarized signal will be extinguished. Thus, the disk inclination can be 
derived from spectro-polarimetric observation, also in cases where this is not 
possible in unpolarized light. 

In this section, the effect of polarization on SEDs is shown for an illustrating toy model as well as for a model of a protoplanetary disk.

\begin{figure}
\includegraphics[width=0.7\linewidth]{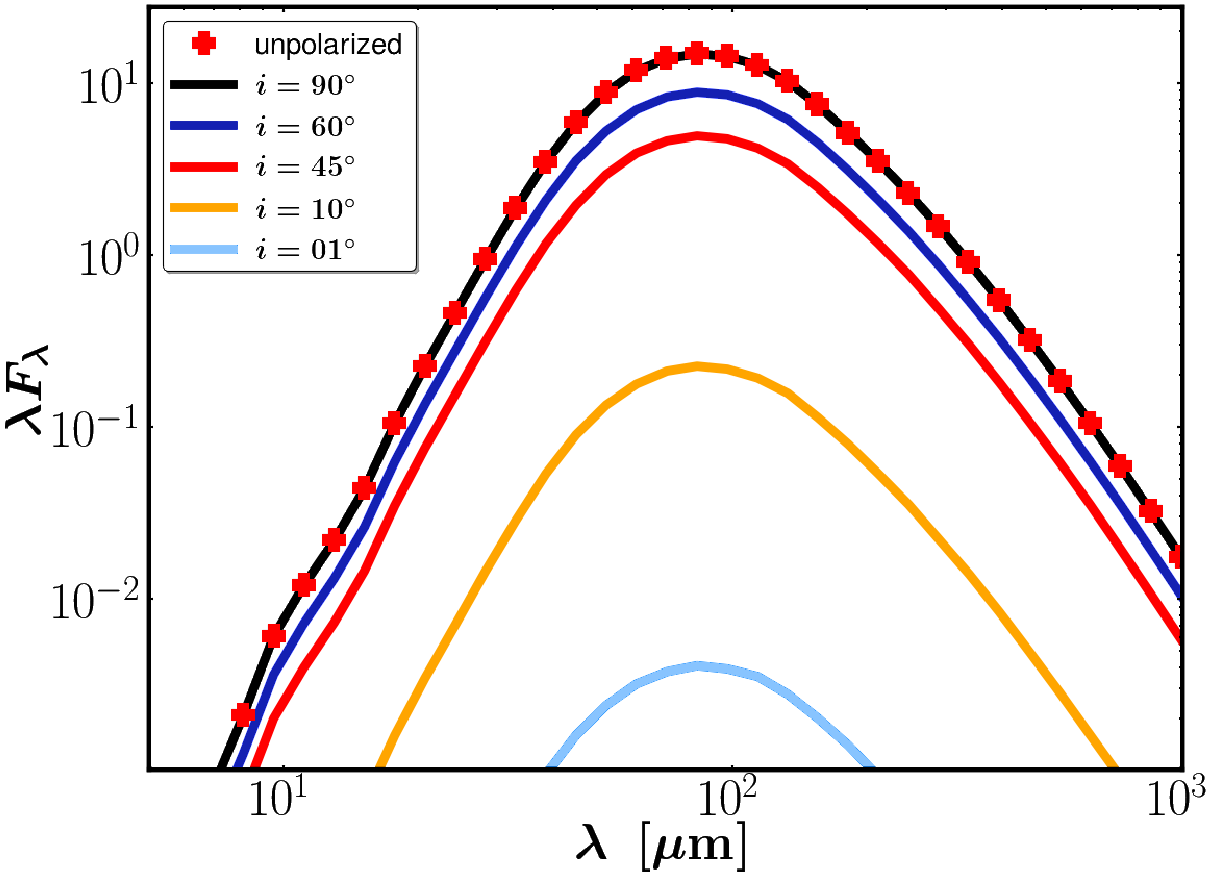}\hspace*{0ex}
\includegraphics[width=0.3\linewidth]{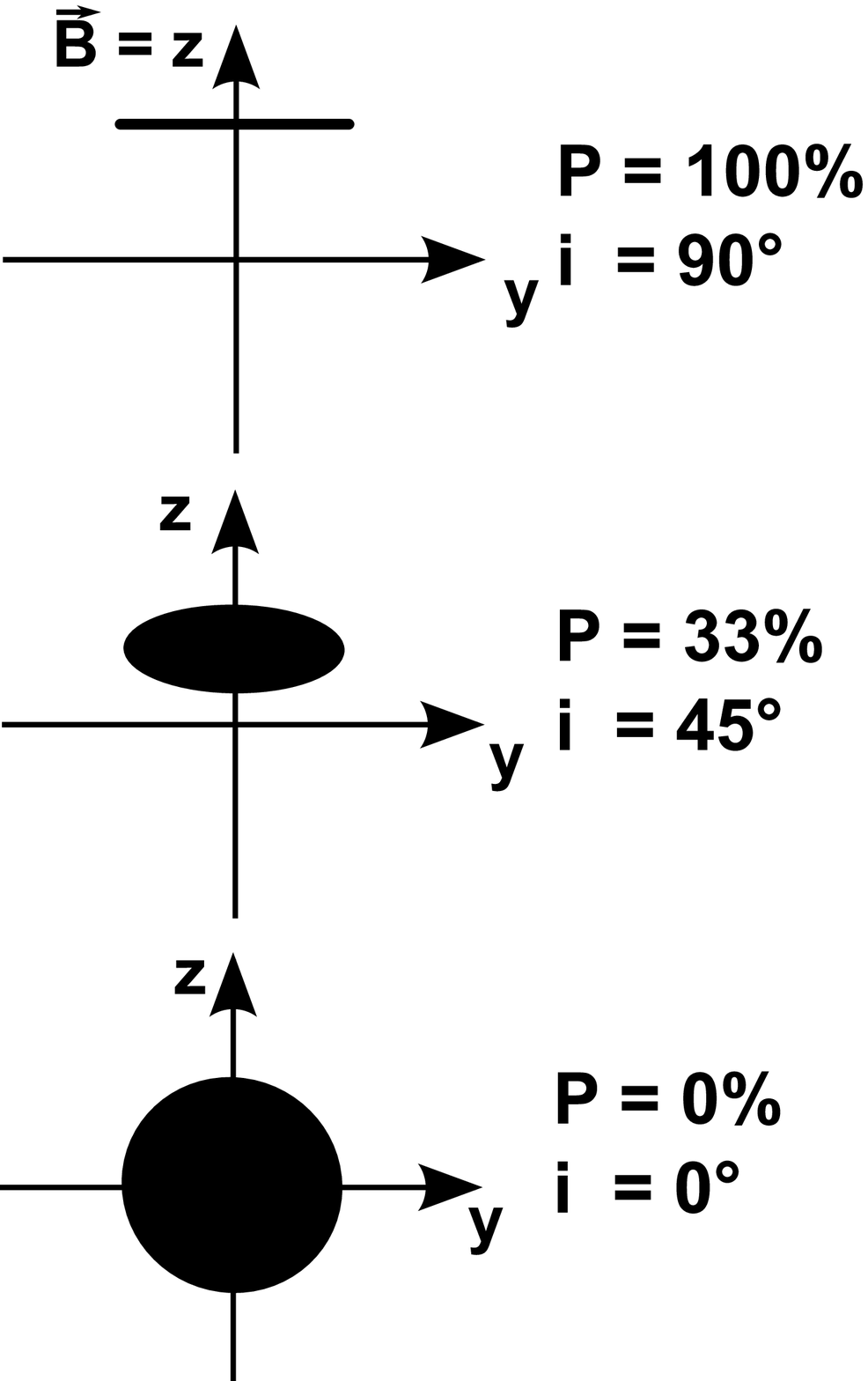}
\caption[\textsc{Spectro-polarimetry of a sphere} containing a homogenous 
magnetic field.]{\textsc{Spectro-polarimetry of a sphere} containing a 
homogenous magnetic field. The effect of inclination is clearly identifiable in 
the SED of the polarized emission. At an inclination $i=90\degree$ the observer 
looks directly at the edge of all dust grains, thus the emission increases to 
its maximum ($P=100\%$, assuming razor-thin dust grains). By changing 
inclination towards a face-on observation of the dust grains, the
polarization degree drops to $P=33\%$ at $i=45\degree$ and becomes $P=0\%$ at 
the face-on position where $i=0\degree$.}\label{pic:sphere_sed}
\end{figure}\vspace*{2ex}

{\em The illustrating toy model: A sphere}\label{subsec:toymodelsed}
\mbox{ }\\[0.5ex]
The most simple case for spectro-polarimetry is a spherical dust distribution, $\rho_{\rm dust}\left(r\right)$, 
containing a homogenous magnetic field $\mathbf{B}$,
\begin{equation}
 \rho_{\rm dust}\left(r\right) \sim r^{-1},\qquad \mathbf{B} \sim \mathbf{e}_{\rm z},
\end{equation}

\noindent where $\mathbf{e}_{\rm z}$ is the unit vector in z-direction. To 
restrict the ana\-lysis only to effects of inclination, all dust grains are 
assumed to be perfectly aligned to the magnetic field. The total 
(unpolarized) emission of this spherical symmetric model is independent of the 
inclination (Fig.~\ref{pic:sphere_sed}). However, by changing the inclination from an edge-on position to a 
face-on position relative to the aspherical dust grain, the polarization 
degree changes. The polarization degree is maximal ($P=100\%$ in the case of a 
razor-thin grain) at the edge-on position ($i=90\degree$), drops to $P=33\%$ at 
an inclination of $i=45\degree$, and at the face-on position ($i=0\degree$) the 
polarization degree disappears ($P=0\%$; Fig.~\ref{pic:sphere_sed}). Thus, by determining the SED of the polarized dust emission relative to the unpolarized radiation, the inclination of the 
object can be derived from the SED. A model of a protoplanetary disk is 
discussed in the next section.

\begin{figure}
\includegraphics[width=\columnwidth]{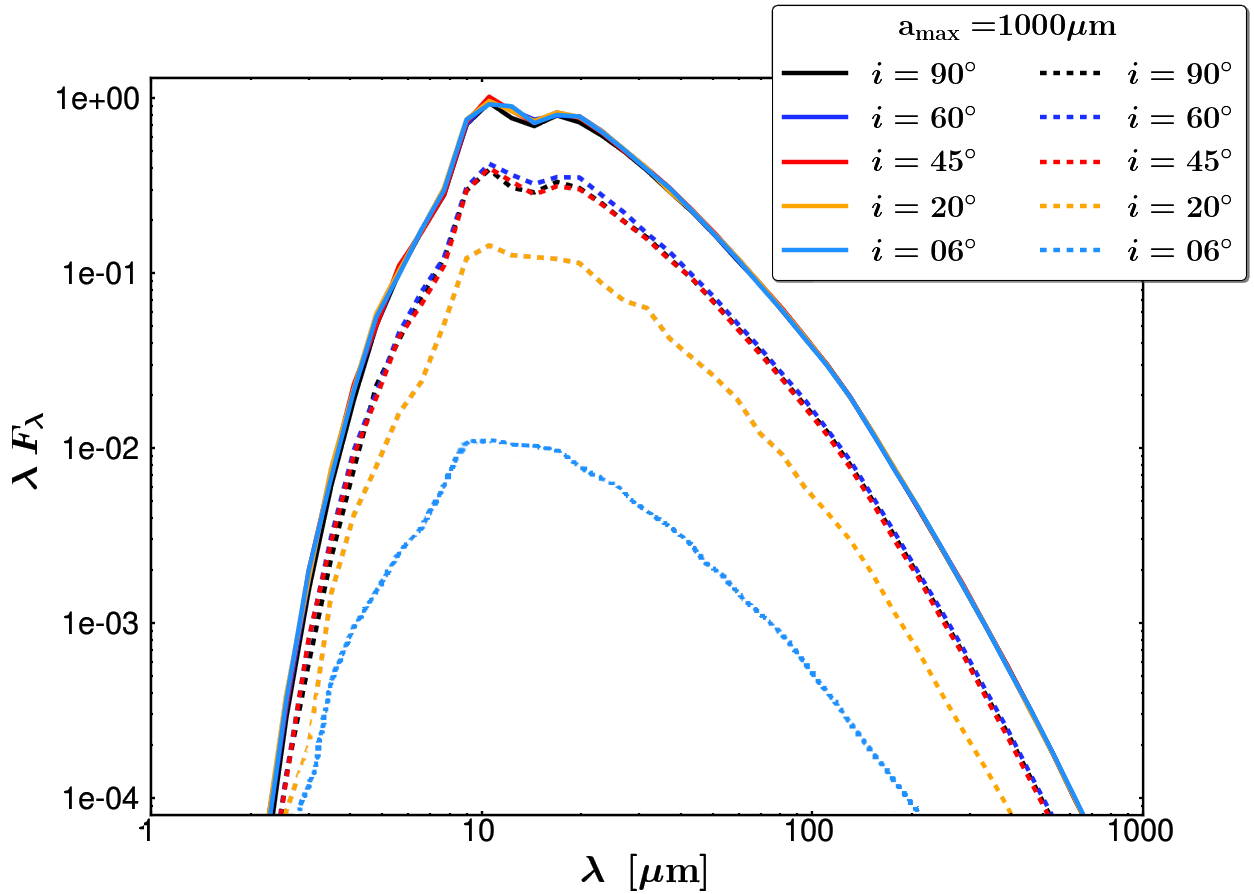}
\caption[\textsc{Spectro-polarimetry of a protoplanetary disk} containing a 
toroidal magnetic field.]{\textsc{Spectro-polarimetry of a protoplanetary disk} 
containing a toroidal magnetic field, computed for a maximal dust grain size of 
$1000\um$. The total (unpolarized) emission is shown in \emph{solid lines}, the 
polarized emission in \emph{dashed lines} both at different inclinations $i$, 
the intensity ($\lambda F_{\rm \lambda}$) is in arbitrary units. Note that the 
solid lines are overlapping each other.}\label{pic:disk_sed}
\end{figure}

\vspace*{2ex}
{\em The protoplanetary disk}\label{subsec:disksed}
\mbox{ }\\[0.5ex]
SEDs of the polarized emission of aspherical dust 
grains in a protoplanetary disk have been computed based on dust grain 
alignment simulations. In Fig.~\ref{pic:disk_sed}, the SEDs are shown for the 
case of large grains with $a_{\rm 
max}=1000\um$. The characteristic $10\um$ silicate feature is clearly seen 
in all SEDs. The disk is optically thin, thus its SEDs at different inclination angles are 
indifferent in total emission. It is prominent that the SEDs in polarized light 
are indistinguishable for high inclinations ($i \gtrsim 45\degree$). In these cases, the polarization pattern observed at high inclinations is dominated by parallel polarization vectors (Fig.~\ref{pic:pol_comp}). A significant drop in 
the amplitude of the SEDs is found for lower inclinations ($i < 45\degree$). As 
it is expected from the radial symmetric polarization pattern corresponding to a
toroidal magnetic field structure, the signal tends to vanish with very small 
inclinations.

As shown here, the inclination of the disk affect the SED in polarized light 
additionally to the common physical parameters derived from SEDs, such as chemical dust composition or dust grain size. These simulations show  that it is possible to distinguish bet\-ween 
different models of protoplanetary disks and magnetic fields therein by taking 
into account not only the SED in the unpolarized light but also the 
polarization information that is carried by the radiation.

\subsection{Spatially resolved polarization maps.} 
\label{subsec:polmaps_resolved}

Previous observations performed with SMA and CARMA traced the polarization information sparsely with low spatial resolution \citep[e.g.,][]{2014Natur.514..597S} or resulted in non-detections \citep{2009ApJ...704.1204H, 2011ApJ...727...85H}. Due to the clear discrepancy between those observations and theoretical models \citep[e.g., \CL;][and this work]{2017MNRAS.464L..61B}, it was suggested to perform more sensitive observations of the dust polarization \citep{2014Natur.514..597S}.
With the Atacama Large (Sub)Millimeter Array (ALMA) it is now possible to spatially resolve protoplanetary 
disks and measure their polarized dust emission on a highly resolved level for 
the first time \citep{2016arXiv161006318K}.

Thus, spatially resolved radiative transfer simulations of the 
polarized dust emission are necessary to analyze such observations. 

\begin{figure*}
\includegraphics[width=\linewidth, keepaspectratio]{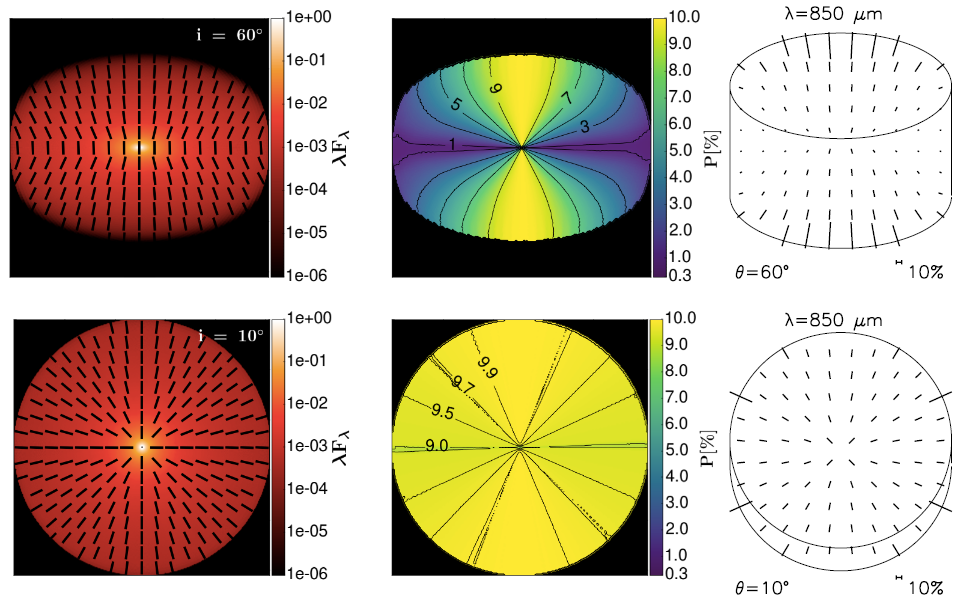}
\caption[\textsc{Polarization maps}, polarizations vectors with underlying total
intensity map.]{\textsc{Polarization maps},  \emph{left:} this work, 
showing total intensities in arbitrary units over-plotted by polarization 
vectors of unit length at a wavelength of $850\um$ and disk inclinations of $i=60\degree$ 
($a_{\rm max}=1000\um$, \emph{top left}) and $i=10\degree$ (\emph{bottom 
left}).   \emph{Center:}  Corresponding maps of the polarization degree, $P$, over-plotted by contour lines of $P$. Values of $P$ below the detection limit of ALMA ($P=0.3\%$) are marked in black. \emph{Right:} Thermal polarization maps from \CL~ (reproduced with 
permission by AAS) at a wavelength of $850\um$ of the emission of disk interior 
+ surface layer at two disk inclinations $\theta = 60\degree$ (\emph{top left}) 
and $\theta = 10\degree$ (\emph{bottom left}). 
}\label{pic:pol_comp}\vspace*{-2ex}
\end{figure*}

Figure~\ref{pic:pol_comp} shows simulations of spatially resolved 
observations of  intrinsically polarized radiation of a protoplanetary 
disk compared to the maps shown in \CL. The 
underlying magnetic field structure is toroidal and the
dust grain alignment is computed as described in Eq.~(\ref{eq:implemented}).
All maps show simulated observations at a wavelength of 
$850\um$ at two different disk inclinations of $10\degree$ and $60\degree$. 
The maps display the characteristic polarization pattern of an 
underlying toroidal magnetic field structure. At a nearly face-on inclination 
of $i=10\degree$, the radially symmetric pattern is clearly visible in the 
results of both codes and the polarization degree is nearly constant as 
expected. Only four vectors in the maps of \CL~ show an increase in polarization 
degree what is potentially caused by boundary effects. The polarization angle, 
the vector orientation (Figure~\ref{pic:pol_comp}, left column), indicates a homogenous field structure but the 
polarization degree (Figure~\ref{pic:pol_comp}, center column) implies a twist in the orientation of 
aligned dust grains caused by the toroidal field structure. However, without the 
knowledge of the underlying magnetic field structure, the drop in the 
polarization degree may be as well interpreted as a lack of order in the 
magnetic field, and thus, an insufficiently strong magnetic field 
influence in this region.
 
\begin{figure*}
\includegraphics[width=\linewidth, keepaspectratio]{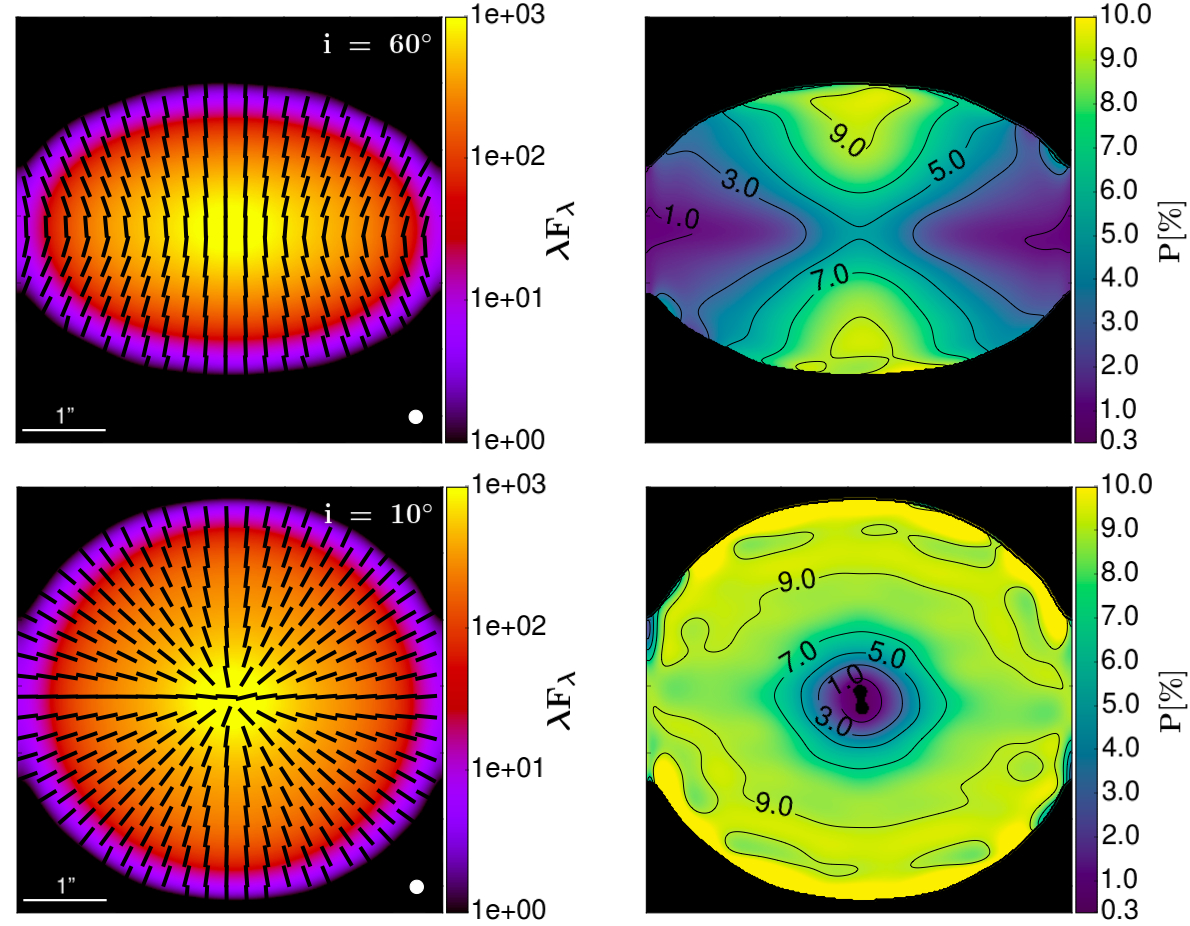}
\caption{\textsc{Simulated ALMA observations} (CASA v4.5.2, band 6, configuration C40-5, spatial resolution of 0.23", including thermal noise) of the intrinsically polarized dust emission at an two different inclinations $i$. {\it Left:} Total intensity map in arbitrary units over-plotted by polarization vectors of unit length indicating the polarization angle only. The polarization vectors are binned over one beam (indicated by the white ellipse). {\it Right:} Corresponding maps of the polarization degree, $P$, over-plotted by contour lines of $P$. Values of $P$ below the detection limit of ALMA ($P=0.3\%$) are marked in black. Note that,in the case of a (almost) face-on disk, the observed Stokes parameters Q and U blur in the center of the disk, resulting in a nulling of the polarized signal.}
\label{pic:pol_ALMA}\vspace*{-2ex}
\end{figure*}

In the more inclined maps 
($i=60\degree$), the radial polarization pattern converts into a rather 
hourglass-shaped structure which results from the projection effect along the 
line-of-sight and is found in both maps. The same also applies to the degree of 
polarization. Close to the vertical symmetry axis, the polarization signal is 
the strongest and drops only slowly towards the center of the disk. Along the 
horizontal symmetry axis of the disk, the polarization signal is canceled out 
almost completely due to projection effects. Only with increasing distance to 
this axis, the  polarization signal becomes stronger again. This behaviour is 
found in both studies. The polarization 
pattern shows the characteristic radial polarization structure of a purely 
toroidal magnetic field as well as the drop of the polarization degree towards 
the center of the disk. The 
observations of face-on oriented disks are the ideal case for
identifying this radially symmetric magnetic field structure. It is obvious that 
spatially unresolved observations of such structures would lead to a 
non-detection because of the radial symmetry of the polarization pattern. In 
unresolved face-on observations of a protoplanetary disk, the polarization
states of the waves emitted from different points in that disk will cancel each 
other out. Therefore, the spatial resolution is essential for polarization 
observations of protoplanetary disks performed with the polarization modes of, 
e.g., ALMA. These new simulations of the polarized thermal dust emission can be directly compared to future ALMA observations by applying the {\sl Common Astronomy Software Applications package} (CASA $v\,4.5.2$; Fig.~\ref{pic:pol_ALMA}). In the more inclined map ($i=60\degree$), we find the same behaviour of the polarization vectors in the simulated ALMA observations as in the pure radiative transfer simulations. However, in the rather face-on map ($i=10\degree$), we find a polarization nulling in the center of the disk caused by blurring of Stokes Q and U in the simulated observation.

The tool presented in this study can be used to derive physical parameters
from observations of polarized dust emission, or to predict observational 
results, by adaption of the parameters of the underlying disk, dust, and 
magnetic field model. 

It is interesting to compare the instrinsic thermal polarization to the effect of self-scattering. 
In recently published calculations on the effect of self-scattered thermal emission spherical dust grains were assumed \citep{2015ApJ...809...78K, 2016arXiv161006318K, 2016MNRAS.456.2794Y}. These models show an upper limit to the polarization degree of $\sim 3\%$. Moreover, they show that the expected polarization structure created by self-scattering has circular symmetry, while deviations can occur with changing optical depth. As pointed out in this work, the polarization degree resulting from the intrinsical thermal polarization depends not only on the relativ orientation of grains along the line-of-sight. Instead, the maximal polarization degree also depends on the axis ratio of the grains. The grain axis ratio, on the other hand, is a free parameter. Thus, the intrinsic polarization does not show a strong upper limit in the degree of polarization such as self-scattering. Polarization degrees of $\sim 10\%$, which exceed the upper limit of self-scattering polarization, as observed in the protoplanetary disk HD142527 \citep{2016arXiv161006318K} can be explained by the instrinsic thermal polarization of dust grains with axis ratios of 1:1.3 (see Figs.~\ref{pic:pol_comp},~\ref{pic:pol_ALMA}).

\section{Conclusions} \label{conc}

The 3D~radiative transfer study that has been presented in this work considers 
dust grain alignment and polarized dust emission of aspherical grains. This tool enables the comprehensive analysis of (sub-)mm observations of polarized radiation of protoplanetary disks performed with, e.g., ALMA. The major 
features, implemented in the context of this work, are the following.

\begin{enumerate}
 \item This study makes use of a hybrid approach to solve the scattering problem (i.e., scattering by aspherical grains) that 
combines treatment of spherical and aspherical grain shape models.

 \item The anisotropy of the radiation field, as well as the efficiency of dust 
grain alignment is computed for a given
dust and disk model.

\item We find that the alignment efficiency exceeds the threshold of $10$ throughout our models of typical protoplanetary disks. This implies that grain alignment in disks is expected to be a rather common phenomenon. However, a broader parameter study is needed in order to strengthen this finding.

 \item SEDs of the polarized emission of aligned aspherical dust grains are 
a tool to distinguish between different disk
models and magnetic fields therein, where SEDs of the total (unpolarized) 
emission are indistinguishable.

 \item  Spatially resolved polarization maps trace the intrinsically polarized dust emission 
of protoplanetary disks depending on dust and
disk parameters. These simulated polarization maps enable the preparation
and analysis of observations of polarized dust emission performed with, e.g., ALMA.
\end{enumerate}

\noindent The unprecedented sensitivity and resolution of ALMA finally allow 
for spatially observations of the polarized dust emission of protoplanetary 
disks. The polarization signal which was previously canceled out by a lack of 
sufficient spatial resolution, is now observable. Such observations need to be analysed by
radiative transfer simulations as those presented in this work. In this way, differences in
underlying dust, disk, and magnetic field models can be 
revealed. Most recently, \cite[][]{2017arXiv170102063T} emphasized that magnetic fields might be preferentially detectable through mid-infrared polarimetric observations. The applicability of our analysis tools to this wavelength range will be considered in a future study.
Studies such as the presented work are essential to finally observationally constrain the magnetic field influence on the evolution of protoplanetary disks.

\section*{Acknowledgements}

GHMB gratefully acknowledges financial support by the DFG under contract WO857/11-1 within the frame of the DFG Priority Program1573: The Physics of the Interstellar Medium, as well as by the Millennium  Science Initiative (Chilean  Ministry  of  Economy),  through 
grant Nucleus RC13007. 



\bibliographystyle{mnras}
\bibliography{biblio} 


\bsp	
\label{lastpage}
\end{document}